\documentstyle[epsfig,12pt]{article}
\catcode`\@=11 
\let\rel@x=\relax
\newcount\timecount
\newcount\hours \newcount\minutes  \newcount\temp \newcount\pmhours

\hours = \time
\divide\hours by 60
\temp = \hours
\multiply\temp by 60
\minutes = \time
\advance\minutes by -\temp
\def\hour{\the\hours}
\def\minute{\ifnum\minutes<10 0\the\minutes
            \else\the\minutes\fi}
\def\clock{
\ifnum\hours=0 12:\minute\ AM
\else\ifnum\hours<12 \hour:\minute\ AM
      \else\ifnum\hours=12 12:\minute\ PM
            \else\ifnum\hours>12
                 \pmhours=\hours
                 \advance\pmhours by -12
                 \the\pmhours:\minute\ PM
                 \fi
            \fi
      \fi
\fi
}

\def\monthname{\rel@x\ifcase\month 0/\or January\or February\or
   March\or April\or May\or June\or July\or August\or September\or
   October\or November\or December\else\number\month/\fi}

\def\bold#1{\setbox0=\hbox{$#1$}%
     \kern-.025em\copy0\kern-\wd0
     \kern.05em\copy0\kern-\wd0
     \kern-.025em\raise.0433em\box0 }

\def\lsim{\mathrel{\mathpalette\@versim<}}
\def\gsim{\mathrel{\mathpalette\@versim>}}
\def\@versim#1#2{\vcenter{\offinterlineskip
        \ialign{$\m@th#1\hfil##\hfil$\crcr#2\crcr\sim\crcr } }}

\newcount\@tempcntc
\def\@citex[#1]#2{\if@filesw\immediate\write\@auxout{\string\citation{#2}}\fi
  \@tempcnta\z@\@tempcntb\m@ne\def\@citea{}\@cite{\@for\@citeb:=#2\do
    {\@ifundefined
       {b@\@citeb}{\@citeo\@tempcntb\m@ne\@citea\def\@citea{,}{\bf ?}\@warning
       {Citation `\@citeb' on page \thepage \space undefined}}%
    {\setbox\z@\hbox{\global\@tempcntc0\csname b@\@citeb\endcsname\relax}%
     \ifnum\@tempcntc=\z@ \@citeo\@tempcntb\m@ne
       \@citea\def\@citea{,}\hbox{\csname b@\@citeb\endcsname}%
     \else
      \advance\@tempcntb\@ne
      \ifnum\@tempcntb=\@tempcntc
      \else\advance\@tempcntb\m@ne\@citeo
      \@tempcnta\@tempcntc\@tempcntb\@tempcntc\fi\fi}}\@citeo}{#1}}
\def\@citeo{\ifnum\@tempcnta>\@tempcntb\else\@citea\def\@citea{,}%
  \ifnum\@tempcnta=\@tempcntb\the\@tempcnta\else
   {\advance\@tempcnta\@ne\ifnum\@tempcnta=\@tempcntb \else \def\@citea{--}\fi
    \advance\@tempcnta\m@ne\the\@tempcnta\@citea\the\@tempcntb}\fi\fi}

\catcode`\@=12 

\begin{document}
\vskip 20pt

\def\beq{\begin{equation}}
\def\eeq{\end{equation}}
\def\bea{\begin{eqnarray}}
\def\eea{\end{eqnarray}}
\def\MSbar {\hbox{$\overline{\hbox{MS}}\,$}}
\def\smallMSbar {\hbox{$\overline{\hbox{{\footnotesize MS}}}\,$}}
\def\aBj{\hbox{$a^*_{Bj}$}}
\def\Obs{\hat{O}}
\def\CI{{\cal C}}
\def\naive{na\"{\i}ve}
\def\etal{{\em et al.}}
\newcommand{\mycomm}[1]{\hfill\break{ \tt===$>$ \bf #1}\hfill\break}
\newcommand{\eqref}[1]{eq.~(\ref{#1})}   
\newcommand{\eqrefA}[1]{(\ref{#1})}   
\def\newline{\hfill\break}
\def\tfrac#1#2{{#1/#2}} 
\def\ra{\hbox{$\rightarrow$}\ }

\def\mpade{{\hbox{\scriptsize Pad\'e}}} 
\def\itemsepreset{\setlength{\itemsep}{0em}}
\def\ds{\displaystyle}
\def\BF{$\beta$ function}
\def\fourloopB{four-loop \BF}
\def\nl{\hfill\break}
\def\ni{\hfill\break$\bullet$ \ }
\def\Ave{\hbox{$\langle A \rangle$}}
\def\Bve{\hbox{$\langle B \rangle$}}

\parskip 0.3cm
\begin{titlepage}
\begin{flushright}
CERN-TH/96-327\\
TAUP-2389-96\\
hep-ph/9612202\\
\end{flushright}

\begin{centering}
{\large{\bf
A Prediction for the 4-Loop $\beta$ Function
in QCD\\}}
\vspace{.4cm}
{\bf John Ellis}\\
\vspace{.05in}
Theoretical Physics Division, CERN, CH-1211 Geneva 23, Switzerland \\
e-mail: john.ellis@cern.ch \\
\vspace{0.4cm}
{\bf Marek Karliner}\\
\vspace{.05in}
School of Physics and Astronomy
\\ Raymond and Beverly Sackler Faculty of Exact Sciences
\\ Tel-Aviv University, 69978 Tel-Aviv, Israel
\\ e-mail: marek@vm.tau.ac.il
\\
\vspace{0.4cm}
and\\
\vspace{0.4cm}
{\bf Mark A. Samuel\footnote{Permanent address:
Department of Physics, Oklahoma State University,
Stillwater, Oklahoma 74078, USA}
}\\
\vspace{.05in}
Department of Physics, McGill University\\
Montreal,P.Q., Canada\\
and\\
Stanford Linear Accelerator Center\\
Stanford University\\
Stanford, CA 94309, USA\\
e-mail: physmas@mvs.ucc.okstate.edu   \\ 
\vspace{0.4cm}
{\bf Abstract} \\
\end{centering}
\bigskip
{\small
We predict that the four-loop contribution $\beta_3$ to the QCD $\beta$
function in the \MSbar \break
prescription \,is\, given \,by
\ \hbox{$ \beta_3 \,\simeq\,
23,600(900)\,-\,6,400(200)\,N_f\,+\,350(70)\,N_f^2\,+\,1.5\,N_f^3$\,,}
\break
where $N_f$ is the number of flavours and the coefficient of $N_f^3$
is an exact result from large-$N_f$ expansion. In the
phenomenologically-interesting case $N_f = 3$, we estimate
$\beta_3 = (7.6 \pm 0.1) \times 10^3$. We discuss our estimates
of the errors in these QCD predictions,
basing them on the demonstrated accuracy of
our method in test applications to the $O(N)$ $\Phi^4$ theory, and
on variations in the details of our estimation method, which
goes beyond conventional Pad\'e approximants by estimating and
correcting for subasymptotic deviations from exact results.

} 
\vfill
\begin{flushleft}
CERN-TH-96/327\\
November 1996 \hfill
\end{flushleft}
\end{titlepage}
\vfill\eject

A perennial problem in Quantum Field Theory is the search for
calculational techniques that go beyond conventional perturbation
theory. These are needed in order to control higher-order terms
in what are in many cases, such as QED and QCD, asymptotic
perturbation series, and to make contact with essentially
non-perturbative methods such as the lattice. To be credible,
any proposed technique for going beyond perturbation theory must
demonstrate its ability 1) to make significant predictions with
2) reliable error estimates. One candidate technique is that of
Pad\'e approximants, which have previously met these criteria
in applications to problems in condensed-matter theory and
statistical mechanics~\cite{padeworks}, in particular.

We have been working to extend these successes to
four-dimensional Quantum Field Theories such as QED and
QCD~\cite{SEK,PBB}.
However, the perturbative calculations
at any given order in these gauge theories
are considerably more complicated than
in many lower-dimensional field theories.
Thus the calculated perturbative QCD series are often much
shorter than those familiar in condensed-matter and
statistical-mechanical applications. This is on the one hand
an opportunity, because it accentuates the need for auxiliary
techniques. However, it is on the other hand a challenge for
the Pad\'e approach, which is based on the available
perturbative results. Shorter series provide less input
into the Pad\'e machinery, increasing unavoidably the errors
and making it difficult to cross check their magnitudes.

A welcome opportunity to test the Pad\'e method is currently
provided by the perturbative series for the renormalization-group
$\beta$ function in QCD~\cite{betaQCD}, which is particularly interesting
for several reasons:
\begin{itemize}
\setlength{\topsep}{0em}
\setlength{\partopsep}{0em}
\setlength{\parskip}{0em}
\setlength{\itemsep}{0.1em}
\item[{}] 
\vskip-2.4em
\item[1)] It is the most
fundamental quantity in perturbative QCD, and its behaviour
impinges on many of the most basic issues in this theory,
such as the question whether the QCD coupling strength $\alpha_s$
approaches a finite value in the infrared limit~\cite{Stevenson}.
\item[2)] It has already been calculated exactly
to three-loop order for any number
of flavours $N_f\,$~\cite{betaQCD}, providing as much information as is
available
from any perturbative QCD series to be incorporated into any attempted
extrapolation.
\item[3)] This and the availability of four- and five-loop
calculations of the $\beta$ function in $O(N)$ scalar field theory
in four dimensions~\cite{vkt,Chetyrkin,Exact5Loop}, which has a similar
structure, facilitate the estimation of plausible errors.
\item[4)] The calculation of the four-loop contribution $\beta_3$ to
$\beta$ seems to be within the scope of available perturbative
techniques,
and we are aware of at least one project \cite{JoosPrivate}
that is under way to
evaluate it, which means that a prediction can be checked in the
foreseeable future.
\end{itemize}

We predict in this paper that
\beq
\beta_3 \simeq 23,600(900)-6,400(200)\,N_f+350(70)\,N_f^2+1.5\,N_f^3
\label{prediction}
\eeq
in the \MSbar prescription,
where the coefficient of the last term is an exact input
calculated using large-$N_f$ techniques~\cite{Gracey}. In the
case $N_f = 3$ of phenomenological interest, we predict
$\beta_3 = (7.6 \pm 0.1) \times 10^3$. The estimates
of the errors in our predictions are discussed in more detail
below. They are based on the demonstrated accuracy of
our method in test applications to the $O(N)$ $\Phi^4$ theory, and
on variations in the details of our estimation method, which
goes beyond conventional Pad\'e approximants by estimating and
correcting for subasymptotic deviations from exact results~\cite{APAP}.

We now review the formalism of Pad\'e approximants~\cite{Pade},
introducing the
notation we use in this paper. Considering a general perturbative series
$ P(x) = \ds\sum_{n=0}^\infty S_n x^n$, the Pad\'e approximant
$P_{[N/M]}(x)$ of order
[N/M] is the ratio of polynomials $A_N(x)$ of order N and
$B_M(x)$ of order M, chosen so that
\begin{equation}
P_{[N/M]} \, = \, {A_N(x) \over B_M(x)} \, = \, P(x) \, + \, {\cal O}(x^{N+M+1})
\label{pnm}
\end{equation}
One may use the $P_{[N/M]}(x)$ in either of two ways:
one may interpret the coefficient $S^\mpade_{N+M+1}$ of $x^{N+M+1}$ in a
power-series expansion of $P_{[N/M]}(x)$ as the Pad\'e
Approximant Prediction (PAP) for the next perturbative coefficient $S_{N+M+1}$,
or one may interpret the full expression $P_{[N/M]}(x)$ as a Pad\'e
Summation (PS) of the entire series $P(x)$~\cite{PBB}. In this paper, we
shall mainly
be concerned with refinements of the PAP's for Quantum Field-Theoretical
perturbation series, but we shall also comment at the end on an
application of PS.

We have exhibited previously~\cite{SEK,PBB} mathematical conditions
sufficient to guarantee that PAP's converge to $S_{N+M+1}$ as $N$ and $M$
are increased. This is in particular true for any series whose perturbative
coefficients have the property that
\begin{equation}
{S_{n+1} S_{n-1} \over S_n^2} \,\simeq\,
\left\{
\begin{array}{cccc}
&&\ds 1 + {A \over n} \\
&\hbox{or} \\
&&\ds 1 + {B \over n^2} \\
\end{array}
\right.
\label{condition}
\end{equation}
as $n \rightarrow \infty$. The first of these conditions is satisfied by
any perturbative series that is dominated by a finite set of renormalon
poles, and the latter condition is satisfied by the Borel transforms of
such series, guaranteeing the utility of PAP's in many applications to
perturbative QCD~\cite{PBB}.

We have, moreover, been able to derive~\cite{PBB} asymptotic expressions
for the corresponding relative errors of PAP's:
\begin{equation}
\delta_{N+M+1} \equiv
{S^\mpade_{N+M+1} - S_{N+M+1}
\over S_{N+M+1}}
\,\simeq\,
\left\{
\begin{array}{c}
\ds {-} {M! A^M \over N^M} \\
\\
\ds{-}\,\frac{M!\,B(B+1)\ldots(B+M-1)}{N^{2M}} \\
\end{array}
\right.
\label{relerror}
\end{equation}
as $N \rightarrow \infty$, for fixed $M$. There is empirical evidence that
the relative errors of PAP's are fitted even better
by expressions of the
form (\ref{relerror}) with the replacements
\begin{equation}
N \rightarrow N + M + aM + b
\label{ab}
\end{equation}
where the parameters $a,b$ are to be fitted on a case-by-case
basis.\footnote{In certain simple cases such amended formulae for the
relative errors (\protect\ref{relerror}) turn out to be {\em exact}.}
See for example the comparison between PAP's and the true values of
perturbative coefficients in QCD for a large number of flavours $N_f
\rightarrow \infty$ shown in~\cite{SEK} and compared with asymptotic
formulae for the relative errors in~\cite{EK}.

These convergence theorems and the successes of the asymptotic error
formulae (\ref{relerror}) lead us to propose that one use the
estimated asymptotic errors to correct PAP's systematically to the
Asymptotic-Pad\'e Approximant Predictions (APAP's):
\begin{equation}
S^{APAP}_{N+M+1} = {S^\mpade_{N+M+1} \over 1 + \delta_{N+M+1}}
\label{APAP}
\end{equation}
We document elsewhere~\cite{APAP} numerical evidence that APAP's
excel even over conventional PAP's in the accuracy with which they
reproduce known coefficients in QCD perturbation series, using as inputs
calculations of lower-order coefficients. Our objective in this paper is
to {\it predict}, with a meaningful and credible error estimate, the
next unknown (four-loop) term in the perturbative series for the
QCD $\beta$ function, whose exact calculation by conventional methods
is currently being undertaken~\cite{JoosPrivate}. If our prediction turns
out to be correct
within the stated errors, it will provide reason to believe Pad\'e
predictions for other coefficients  that are less tractable to
calculate exactly.

We denote the QCD \BF\ by
$\beta(x) = {-} \left( \sum_{n=0}^\infty \beta_n x^{n+1} \right)$,
where $x \equiv \alpha_s/(4\pi)$. The known coefficients for $N_c=3$
in the \MSbar prescription are~\cite{betaQCD}
\ \hbox{$\beta_0 = 11 - (2 N_f / 3)$}, \
\hbox{$\beta_1 = 102 - (38 N_f / 3)$} \ and \
\hbox{$\beta_2 = (2857/2) - (5033/18) N_f + (325/54) N_f^2$}.
It is possible to use the [0/1] PAP to estimate
$\beta_2$ on the basis of the exact values of $\beta_{0,1}$.
This is done by fitting the numerical values of $\beta_2(N_f)$
obtained from the [0/1] PAP for $N_f=0,\ldots,4$
to the polynomial form $\beta_2={\cal A} + {\cal B} N_f + {\cal C} N_f^2$.
Since the leading coefficient of $N_f$ is known at each order in loop
expansion \cite{Gracey}, we anticipate here the approach to be used in
the 4-loop case and include its exact value in the fit,
finding
\beq
\beta_2 = 946- 177 \,N_f + 6.02\,N_f^2
\label{lowestPAP}
\eeq
Although qualitatively correct, this estimate is not very accurate,
which was hardly to be expected. The use we make of the [0/1] PAP
is to estimate, for each individual value of $N_f$, the normalization
coefficient $A$ or $B$ of the asymptotic
correction (\ref{relerror}), which we then use as an input in
calculating the [1/1] APAP (\ref{APAP}).

The result we obtain clearly depends on our choice between the possible
asymptotic forms in (\ref{condition}), which is
$\beta_{n+1} \beta_{n-1} / \beta_n^2 \sim 1 + A/n$.
In contrast to the case with many QCD observables such as
 $\alpha_R /\pi\equiv \left(R_{{}_{e^+e^-}}-\sum_f Q_f^2 \right)$ in
$e^+\,e^- \rightarrow \hbox{hadrons}$,
or the anomalous dimension of the quark mass insertion, we are
unaware of any proof that there is a renormalon: $\beta_n \sim n!$
in the \MSbar prescription used in this paper,
which would be a sufficient condition for this assumption to
apply~\cite{PBB}~\footnote{Consider, however, a generic change
in renormalization prescription from \MSbar to one in which the QCD
correction to an observable (e.g., $\alpha_R$) is treated as an
effective charge. In order to cancel the factorial growth of the \MSbar
perturbative coefficients for the observable, it is clear that the
$\beta$-function coefficients must also diverge factorially.}.
More is known about the $\beta$ function in supersymmetric QCD.
Indeed, it is trivial in $N = 2$\  QCD, being given entirely by
the one-loop term~\cite{N2QCD}. In the case of $N = 1$ \ QCD with matter,
it is known that in a background-field Pauli-Villars
prescription~\cite{N1QCD}
\beq
\beta(g) = - {g^3 \over 16 \pi^2} {3 N_c - N_f + \gamma(g^2)\,N_f
\over 1 - N_c \,(g^2 / 8 \pi^2)}
\label{nequalsone}
\eeq
where $\gamma(g^2)$ is the anomalous dimension of the mass operator. If the
perturbative coefficients in $\gamma(g^2)$ are dominated by a
renormalon, then also $\beta_n \sim n!$, and our asymptotic
assumption applies~\footnote{We note that the positions of
renormalon singularities are not thought to depend on the
renormalization scheme chosen, just their strengths. Therefore,
one may expect this conclusion to carry over to the \MSbar\
prescription used in this paper.}.
On the other hand, if one considers the limit
\hbox{$N_c \,g^2$ fixed} and \hbox{$N_f/N_c = 3 - \epsilon$ fixed}, the
numerator of $\beta(g)$ (\ref{nequalsone}) is determined by the
${\cal O}(g^2)$ term in $\gamma$, and the [1/1] PAP is {\it exact}!

These considerations support our assumption that the leading
behaviours of the perturbative coefficients are as given in the top
line of (\ref{condition}), leading to the corresponding expression
(\ref{relerror}) for the relative error, and hence justifying the
corresponding APAP correction (\ref{APAP}), though they
by no means prove it.

As a cross check on our method, and to aid in assessing its likely errors,
we first apply it to the $O(N)$ $\Phi^4$ theory in $4$ dimensions,
whose \BF\ we denote by $\beta(g) = \sum_{n=0}^\infty \beta_n g^{n+2}$.
These \BF\ coefficients are known for $n \le 4$, and it is known that
$\beta_n \sim n!$ as $n \rightarrow \infty$ in this theory~\cite{Lipatov},
justifying
the APAP procedure with the first form of subasymptotic correction in
(\ref{relerror}). We follow
the same procedure as advocated above for QCD, namely we use the
known results for $\beta_{0,1}$ in [0/1] PAP's for various values of
$N$ and then apply the APAP correction (\ref{APAP}) to the naive [1/1]
PAP~\footnote{This was used as an order-of-magnitude estimate
in~\cite{KS}.}. Our results for $\beta_3$ in
$O(N)$ $\Phi^4$ theory are shown in Fig.~1(a).
We see that the naive [1/1] PAP falls considerably below the
known exact result for $\beta_3$ in this case, which
is known to be a polynomial in $N$:
$\beta_3 = {\cal A} + {\cal B}N +{\cal C}N^2 + {\cal D}N^3$, shown as
the solid line in Fig.~1(a). It is clear from this figure
that the [1/1] APAP with the $A/n$ type of correction (\ref{relerror})
is startlingly accurate, reproducing qualitatively the known polynomial
form and quantitatively the values of $\beta_3$ for $N \le 4$. We
have in fact determined empirically that the most accurate prediction
of the coefficients in the polynomial expansion of $\beta_3$ may be
obtained by a naive average \Ave\ of the correction coefficient $A$ over
the values $N = 0, 1, 2, 3, 4$ studied. Using this \Ave\ prescription,
and then making a polynomial fit to
$\beta_3 = {\cal A} + {\cal B}N +{\cal C}N^2 + {\cal D}N^3$, with the
exactly known value
of the ${\cal O}(N^3)$ coefficient ${\cal D}$ \cite{vkt,Chetyrkin,Exact5Loop},
we find
the numerical values of the polynomial coefficients shown
in Table I, reported together with their errors relative to the known exact
coefficients~\footnote{The errors in the polynomial coefficients are
correlated in the fitting procedure, and our errors for individual values
of $N$ are smaller. In the particular case $N=3$, for
example, we find $\beta_3 = 222.4$, to be compared with the exact result
$\beta_3 = 218.8$, an error of $1.6\%$.}.
For comparison, we also display the relative errors
found if one fits the correction coefficient $A$ individually for each
value of $N_f$, as well as the errors found making the (disfavoured)
$B/n^2$ correction, and the errors of the naive PAP.

\vbox{
\begin{center}
Table I \\
APAP results for the polynomial coefficients of $\beta_3$ in $O(N)$ $\Phi^4$\\
\bigskip
\begin{tabular}{|c|c|c|c|c|c|c|} \hline
polynomial  & exact      & APAP        & rel. error & rel. error
& rel. err.                 & rel. err. \\
coefficient & calculation& prediction  & APAP       & APAP,
& APAP,                     & PAP       \\
            &            &             & \Ave\      & $A(N_f)$
&  $\langle B \rangle$      &           \\
\hline\hline
${\cal A}$         & ${-}100.46$     & ${-}104.32$  &  ${+}3.84\%$ &
${+}4.69\%$
& ${-}17\%$ & ${-}31\%$ \\ \hline
${\cal B}$         &  ${-}33.28$     &  ${-}34.39$  &  ${+}3.34\%$ &
${+}1.38\%$
& ${-}17\%$ & ${-}31\%$ \\ \hline
${\cal C}$         &   ${-}2.06$     &   ${-}1.67$  &  ${-}18.87\%$ &
${-}17.18\%$
& ${-}35\%$ & ${-}45\%$ \\ \hline
${\cal D}$         & $6.4\times 10^{-4}$ & input & $\phantom{A^{B^C}}$ &
& & \\ \hline \hline
\end{tabular}
\end{center}
} 

We note that the APAP's with the $B/n^2$ form of correction,
also shown in Fig.~1(a), are considerably less accurate than those
with the $A/n$ form, in accord with what we expect from the
asymptotic results of~\cite{Lipatov}. We also note that the naive
[0/1] PAP is a significantly worse approximation to $\beta_2$ in
$O(N)$ $\Phi^4$ theory than the corresponding [0/1] PAP in QCD.
This implies that the APAP correction is relatively larger in this case,
and leads us to suspect that the [1/1] APAP prediction in QCD might even be
more accurate than that in Fig.~1(a) or Table I.

We now make our prediction for the 4-loop \BF\ coefficient $\beta_3$
in QCD. As in the previous $O(N)$ $\Phi^4$ case,
we first calculate the [1/1] PAP's for $0 \le N_f \le
4$, and then make the appropriate APAP correction assuming that
$(\beta_{n+1} \beta_{n-1}) / \beta_n^2 \simeq 1 + A/n$,
obtaining the predictions shown as crosses in Fig.~1(b). These points
are obtained by
averaging the value of the asymptotic coefficient $A$ over the
values of $N_f$ chosen, which is known to give accurate
predictions in the $O(N)$ $\Phi^4$ case, as we
have already commented in connection with Fig.~1(a). In the
particular case $N_f=3$ of phenomenological interest, we predict
$\beta_3 = (7.6 \pm 0.1) \times 10^3$, with an error which we estimate
conservatively from the corresponding value of $N$ in the
$\Phi^4$ model, and for $N_f = 1,2,4$ we find
$\beta_3 = (17.6, 12.3, 3.8) \times 10^3$ with similar fractional
errors. We also fit the APAP's to the expected polynomial dependence of
form
\hbox{$\beta_3 = {\cal A} + {\cal B}N_f + {\cal C}N_f^2 + {\cal D}N_f^3$},
where the
${\cal O}(N_f^3)$ coefficient ${\cal D}$ is known exactly \cite{Gracey},
obtaining the predictions
shown in Table II.

\vbox{
\begin{center}
Table II \\
$\phantom{a}$\\
APAP results for the polynomial coefficients of $\beta_3$ in QCD\\
\bigskip
\begin{tabular}{|c|c|c|c|c|} \hline
coefficient & APAP, \Ave\ & error     & APAP, $A(N_f)$ & PAP    \\
            & prediction  & estimate  & prediction     & prediction \\
\hline\hline
${\cal A}$         & $23,557$    & $\pm 900$ & $24,086$       & $20,015$
\\ \hline
${\cal B}$         & ${-}6,353$   & $\pm 200$  & ${-}6,572$     &
${-}5,396$ \\ \hline
${\cal C}$         & $346$       & $\pm 70$  & $357$          & $292$
\\ \hline
${\cal D}$         & $1.5$ (input) &  --     &    --          &   --   \\
\hline
\hline
\end{tabular}
\end{center}
} 

We believe on the basis of our $O(N)$ $\Phi^4$ experience that
the most reliable predictions are provided by the \Ave\ APAP's
in Fig.1(b), with the relative errors of the corresponding
$O(N)$ $\Phi^4$ coefficient predictions providing a conservative
estimate of the errors in the QCD case. These errors, which we expect
to be highly correlated as in the $\Phi^4$ case, are also shown in Table
II, as well as the APAP's obtained using $A(N_f)$. We
prefer the \Ave\ APAP's, but are pleased to note that the differences
are comparable with our quoted error estimates.
As an exercise, we have also computed and shown in Fig.~1(b)
as open circles
the APAP results obtained using the $B/n^2$ form of correction,
and also display the naive PAP's shown as diamonds.
We believe that the differences from the $B/n^2$ APAP's are likely
to be overestimates of the errors: as we have discussed above,
the $B/n^2$ APAP's have less theoretical motivation. Finally,
Table II also displays the numerical differences between
our preferred APAP's and the naive PAP's.

As already mentioned, we are aware of a project underway to
calculate $\beta_3$ exactly in QCD, which will enable our
predictions to be tested. If the exact results are consistent
with the predictions of our APAP method, we will then be able to
use them to predict the 5-loop coefficient $\beta_4$ in QCD.
Fig.~2 shows the corresponding \Ave\ and \Bve\ predictions for $\beta_4$
in the
$O(N)$ $\Phi^4$ model, compared with the known exact result in this
model~\footnote{We note in passing that our APAP approach
easily revealed the sign misprint
that appeared in the original version of~\cite{Chetyrkin}, and, even
after sign correction, agrees better with
the corrected value~\cite{Exact5Loop} than with the previous
value~\cite{Chetyrkin} that had a small
numerical error. We
recall that PAP's have on previous occasions shown
that they can identify errors and misprints in tabulations~\cite{padeworks}.}.
The corresponding \Ave\ predictions for the polynomial
coefficients of $\beta_4$ for this model are displayed in Table III. They
were obtained by fitting \Ave\ and the combination $a+b$ of
subasymptotic coefficients in (\ref{ab}) to the known exact results
for $\beta_2$ 
and $\beta_3$ in the $O(N)$ $\Phi^4$ model. Also shown in Table III are
the known exact results and the relative errors.
We see in Fig.~2 that the \Ave\ and \Bve\ predictions are closer in
this higher-order case, as one might have expected.
For comparison,
we recall the prediction~\cite{Chetyrkin} that $\beta_4 = 1405 \pm 80$,
which is to be compared with the exact result $\beta_4 =
1424.29$~\cite{Exact5Loop} and
our prediction that $\beta_4 = 1432$, corresponding to an error of
$0.56 \%$. We would hope that predictions of comparable accuracy
will be possible
in QCD, once the exact result for $\beta_3$ is available.
For completeness, we also provide our APAP
prediction for the 6-loop coefficient $\beta_5$ in the $O(N)$ model,
which has not yet been computed exactly: for $N = 0,1,2,3,4$, we predict
$-\beta_5=$ 11828, 17687, 24958, 33802, 44330, respectively.
For comparison, we recall the prediction~\cite{Chetyrkin}
that $-\beta_5 = 17,200 \pm 50$ in $\Phi^4$ theory with $N=1$.

\vbox{
\begin{center}
Table III \\
$\phantom{a}$\\
APAP results for the polynomial coefficients of $\beta_4$ in $O(N)$ $\Phi^4$\\
\bigskip
\begin{tabular}{|c|c|c|c|} \hline
coefficient &  exact     & APAP &        rel. error\\
            &            & prediction         &               \\ \hline\hline
${\cal A}$         & $1002.0$      & $1002.26$       &  $0.03\%$   \\
\hline
${\cal B}$         & $385.6$    & $390.61$      &  $1.3\%$  \\ \hline
${\cal C}$         & $36.12$    & $38.82$      &  $7.5\%$       \\ \hline
${\cal D}$         & $0.576$    & $0.627$      &  $8.8\%$       \\ \hline
${\cal E}$         & ${-}0.0013$ & input   &               \\ \hline
\hline
\end{tabular}
\end{center}
} 

It is not the purpose of this paper to discuss in detail the
possible phenomenological implications of our prediction for
$\beta_3$ in QCD~\cite{others}. We limit ourselves here to advertizing
the fact that the conventional PS procedure, applied to our APAP
result for $N_f$, predicts the existence of a zero in the
QCD $\beta$ function, and hence an infrared fixed point in the
\MSbar coupling, whose location coincides with that predicted
previously by Stevenson~\cite{Stevenson} using very different arguments.
\bigskip
\begin{flushleft}
{\Large\bf Acknowledgements}
\end{flushleft}

This research was supported by the Israel Science Foundation
administered by the Israel Academy of Sciences and Humanities,
and by a Grant from the G.I.F., the
German-Israeli Foundation for Scientific Research and
Development.
This work was also supported by the US Department of Energy 
under contract No. DE-AC03-76F00515 and grant No. DE-FG05-84ER40215.

\noindent
{\bf Note added in proof:}

After this paper was submitted, T.~van Ritbergen, J.A.M. Vermaseren,
and  S.A.~Larin presented the results of an exact calculation of the
four-loop QCD $\beta$ function (hep-ph/9701390).
The result in their eq.~(8) contains qualitatively new color factors,
corresponding to quartic Casimirs,
which are not present at \hbox{1-,}
2- and 3-loop level, and therefore cannot be
estimated using the Pad\'e method. Consequently, 
our predictions should be compared with the rest of their exact
expression, as shown below.
Writing \
$\beta_3={\cal A}+{\cal B} N_F+{\cal C } N_F^2 + {\cal D} N_F^3$,
\  we have:
\bigskip
\vbox{
\begin{center}
\bigskip
\begin{tabular}{|c|c|c|c|c|} \hline
coefficient & APAP estimate & exact     & relative & number of    \\
            & (uncertainty $\sigma$) & result & error & $\sigma$'s \\
\hline\hline
${\cal A}$ & $23,557(900)$ & 24,633      & ${-}4.4\%$  & 1.20 \\ \hline
${\cal B}$ & ${-}6,353(200)$ & ${-}6375$ & ${-}0.35\%$ & 0.11 \\ \hline
${\cal C}$ & $346(70)$       & 398.5     & ${-}13.2\%$ & 0.75 \\ \hline
${\cal D}$ & $1.5$           &  1.5      & input       & ---  \\
\hline
\hline
\end{tabular}
\end{center}
} 

\def\PL{{\em Phys. Lett.\ }}
\def\NP{{\em Nucl. Phys.\ }}
\def\PR{{\em Phys. Rev.\ }}
\def\PRL{{\em Phys. Rev. Lett.\ }}


\vfill\eject
\def\thefigure{1}
\begin{figure}[H]
\begin{center}
$\phantom{a}$
\vskip-1cm
\mbox{
\epsfig{file=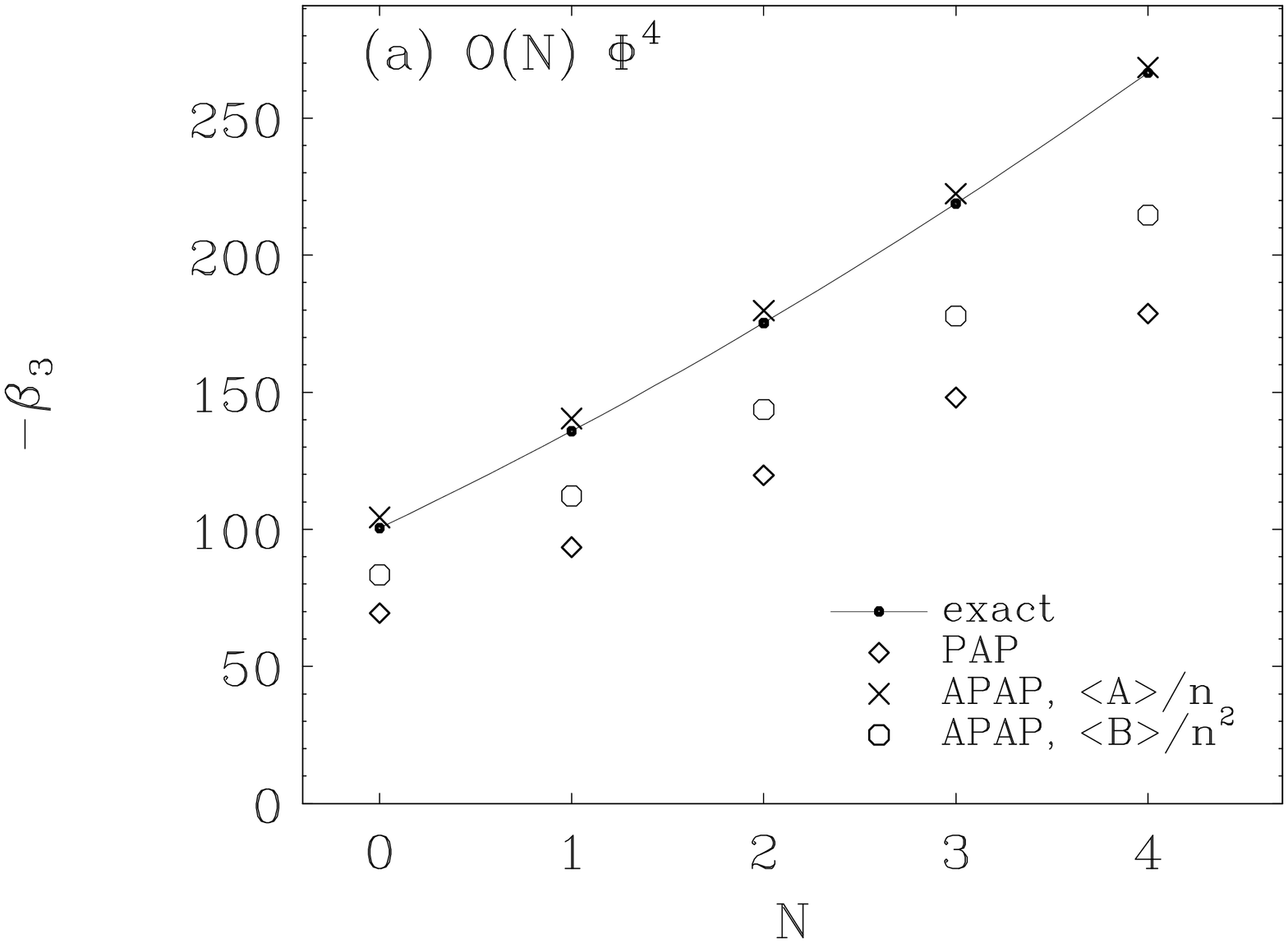,width=8.9truecm,angle=90}}
\\
$\phantom{a}$\\
$\phantom{a}$\\
\mbox{
\epsfig{file=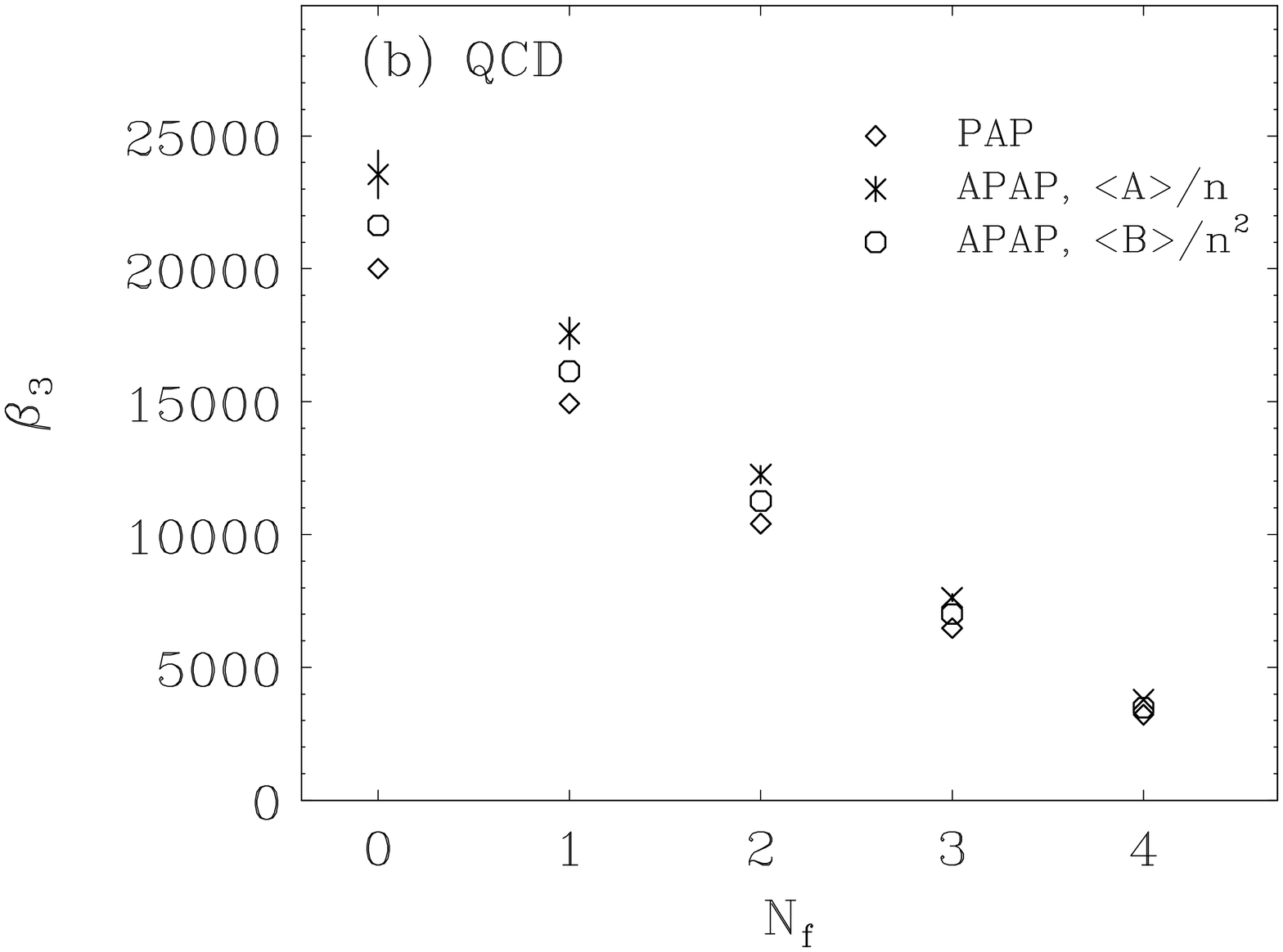,width=8.9truecm,angle=90}}
\caption{\protect\small
(a) The 4-loop $\beta$-function coefficient $\beta_3$ in
$\Phi^4$ theory with $O(N)$ symmetry.
The exact results are denoted by black dots, joined by 
a solid line to guide the eye. Naive PAP results are denoted by  diamonds,
and APAP results obtained from the $A/n$ type of correction are denoted
by crosses. For comparison, also shown are
APAP results obtained from the $B/n^2$ type of correction, denoted
by open circles.
(b)
The 4-loop $\beta$-function coefficient $\beta_3$ in QCD
with $N_f$ flavours, using the same notation as in Fig.~1(a).
The $A/n$ APAP results are assigned error bars obtained
from the relative errors in the corresponding $\Phi^4$
results.}
\label{fig1}
\end{center}
\end{figure}
\vfill\eject

\def\thefigure{2}
\begin{figure}[H]
\begin{center}
\mbox{
\epsfig{file=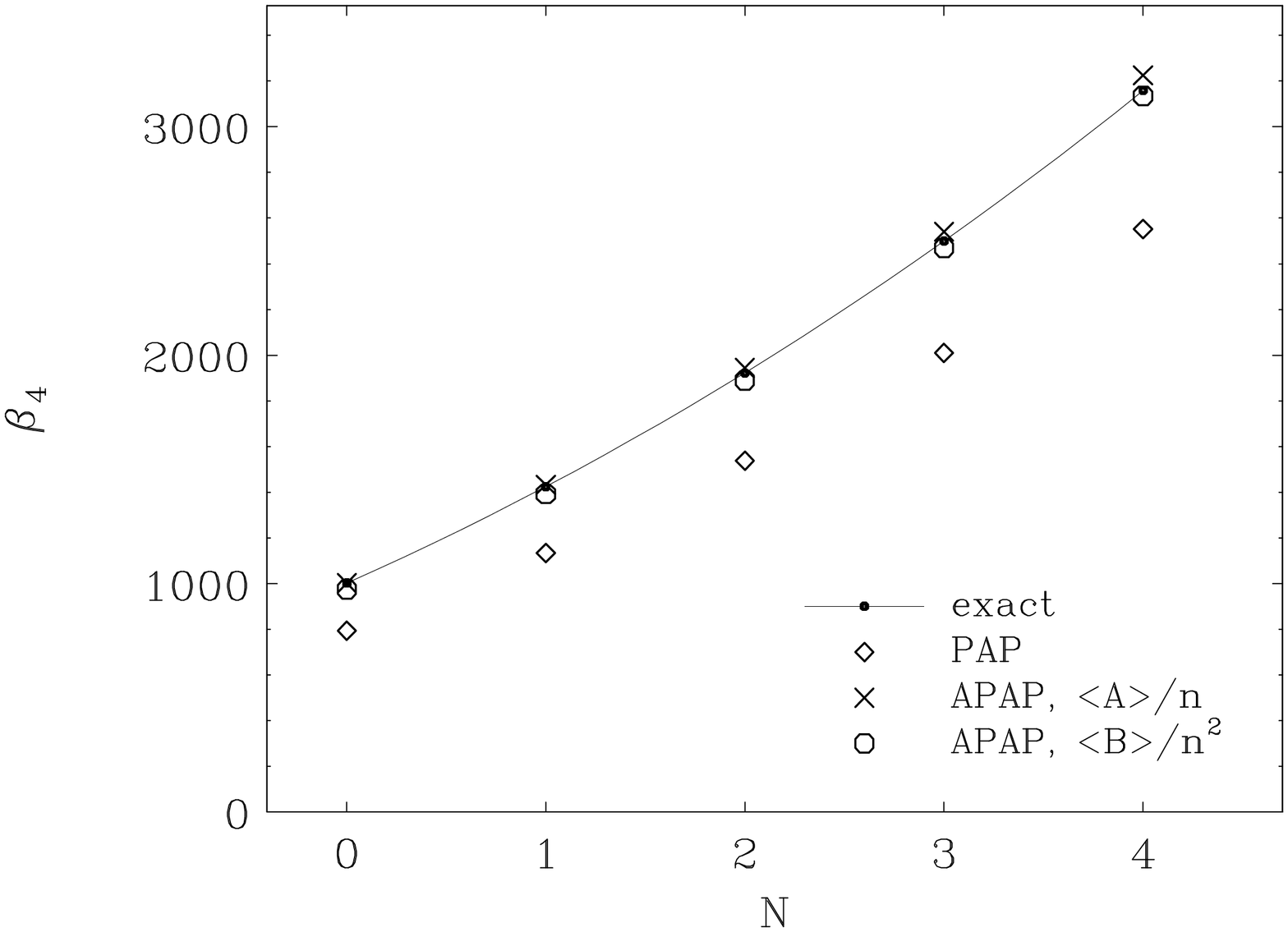,width=9.3truecm,angle=90}}
\caption{\protect\small
The 5-loop $\beta$-function coefficient $\beta_4$ in
$\Phi^4$ theory with $O(N)$ symmetry, using the same notation
as in Fig.~1(a).}
\label{fig2}
\end{center}
\end{figure}
\vskip10cm
\vfill\eject

\begin{thebibliography}{99}

\bibitem{padeworks}
M.A. Samuel, G. Li and E. Steinfelds,
{\it Phys. Rev.} {\bf D48}(1993)869
and {\it Phys. Lett.} {\bf B323}(1994)188;
M.A. Samuel and G. Li, {\it Int. J. Th. Phys..} {\bf 33}(1994)1461 and
{\it Phys. Lett.} {\bf B331}(1994)114.

\bibitem{SEK}
M.A. Samuel, J. Ellis and  M. Karliner,
{\sl Phys. Rev. Lett.} {\bf 74}(1995)4380.

\bibitem{PBB}
 J. Ellis, E. Gardi  M. Karliner and M.A. Samuel,
{\it Phys. Lett.} {\bf B366}(1996)268 and {\it Phys. Rev.} {\bf
D54}(1996)6986.

\bibitem{betaQCD}
O.V. Tarasov, A.A. Vladimirov, A.Yu. Zharkov,
{\it Phys.Lett.} {\bf 93B}(1980)429;
S.~Larin and J.A.M. Vermaseren, {\it Phys. Lett.} {\bf B303}(1993)334,
hep-ph/9302208.


\bibitem{Stevenson} AC. Mattingly and P.M. Stevenson, {\it Phys. Rev.}
{\bf D49}(1994)437;
P.M. Stevenson, {\it Phys. Lett.} {\bf B331}(1994)187.

\bibitem{vkt}
A.A. Vladimirov, D.I. Kazakov and O.V. Tarasov,
{\em Sov. Phys. JETP} {\bf 50}(1979)521 [{\em Zh. Eksp. Th. Fiz.} {\bf 77}
(1979)1035].

\bibitem{Chetyrkin}
K.G. Chetyrkin \etal,
\PL {\bf 132B}(1983)351.

\bibitem{Exact5Loop}
H. Kleinert \etal,
\PL {\bf B272}(1991)39; E -- {\em ibid.} {\bf B319}(1993)545 and
hep-th/9503230.

\bibitem{JoosPrivate} S.~Larin and J.~Vermaseren, private communications.

\bibitem{Gracey} J.A. Gracey,
{\sl Phys. Lett.} {\bf B373}(1996)178.

\bibitem{APAP} J. Ellis, M. Karliner and M.A. Samuel, in preparation.

\bibitem{Pade}
G.A. Baker, Jr.
{\em Essentials of Pad\'e Approximants},
 Academic Press, 1975;
C.M.~Bender and S.A. Orszag,
{\sl Advanced Mathematical Methods for Scientists and Engineers},
McGraw-Hill, 1978.


\bibitem{EK} J. Ellis and M. Karliner, Invited Lectures at the
{\em Int. School of Nucleon Spin Structure}, Erice 1995, CERN preprint
TH/95-334, hep-ph/9601280.

\bibitem{N2QCD}
L.V. Avdeev, O.V. Tarasov,
{\em Phys. Lett.} {\bf 112B}(1982)356;

\bibitem{N1QCD}
V.~Novikov, M.~Shifman, A.~Vainshtein and V.~Zakharov,
{\em Nucl. Phys.} {\bf B229}(1983)381;
M.A.~Shifman and A.I Vainshtein,
{\em Nucl. Phys.} {\bf B277}(1986)456, {\em ibid.} {\bf B359}(1991)571.

\bibitem{Lipatov}
L.N. Lipatov,
{\em Sov. Phys. JETP} {\bf 45}(1977)215 [{\em Zh.Eksp.Teor.Fiz.} {\bf 72}
(1977)411].

\bibitem{KS}
A.L. Kataev and V.V. Starshenko,
{\em Mod. Phys. Lett.} {\bf A10}(1995)235.

\bibitem{others}
J. Ellis, M. Karliner, J. Reid, M.A.~Samuel, E.~Steinfelds, in preparation.

\end{thebibliography}
\end{document}